\pdfoutput=1
\documentclass[11pt]{article}

\usepackage[T1]{fontenc}
\usepackage[utf8]{inputenc}
\usepackage[margin=1in]{geometry}
\usepackage{graphicx}
\usepackage{amsmath,amssymb,amsfonts,amsthm}
\usepackage[numbers,sort&compress]{natbib}
\usepackage[hidelinks]{hyperref}

\DeclareUnicodeCharacter{2011}{\nobreakdash-}
\DeclareUnicodeCharacter{2013}{--}
\DeclareUnicodeCharacter{2019}{'}

\theoremstyle{plain}

\theoremstyle{definition}

\theoremstyle{remark}

\providecommand{\bibcommenthead}{}

\title{Bistability of Exciton--Photon Microcavities in the Ultrastrong-Coupling Regime}
\author{%
Rashed Aljasmi$^{1*}$, Hisham Sati$^{1,2,3}$, and Hichem Eleuch$^{1,4}$\\[0.8em]
\small $^{1*}$Center for Quantum and Topological Systems, NYUAD Research Institute,\\
\small New York University Abu Dhabi, Abu Dhabi, United Arab Emirates.\\
\small $^{2}$Mathematics, Division of Science, New York University Abu Dhabi (NYUAD),\\
\small Abu Dhabi, United Arab Emirates.\\
\small $^{3}$Courant Institute for Mathematical Sciences, New York University,\\
\small New York, 10012, NY, USA.\\
\small $^{4}$Univ. Polytechnique Hauts-de-France, LAMIH, CNRS, UMR 8201,\\
\small F-59313 Valenciennes, France.\\[1em]
\small Contributing authors: \href{mailto:raa10052@nyu.edu}{raa10052@nyu.edu};
}
\date{}

\begin{document}
\maketitle

\begin{abstract}
We investigate a coherently driven exciton--photon microcavity with Kerr nonlinearity in the ultrastrong-coupling regime. When the lower and upper polariton branches are well separated in energy, the full Hopfield--Rabi--Kerr model reduces to an effective single-mode description of the lower polariton. We analyze the stability of the lower-polariton steady states. We show that the resulting bistability is qualitatively similar to that in the strong-coupling regime. However, in the ultrastrong-coupling regime counter-rotating processes and the diamagnetic $A^{2}$ term renormalize the polariton spectrum and composition, changing the effective detuning $\tilde{\Delta}_{1}$ and nonlinearity $U_{\mathrm{LP}}$ $g$-dependency beyond the strong-coupling (RWA) picture. As a result, although the semiclassical bistability criterion keeps its standard Kerr--oscillator form, the turning points and hysteresis window are shifted relative to the strong-coupling prediction.
\end{abstract}


\section{Introduction}\label{sec1}

Semiconductor nanostructures, including quantum wells, two-dimensional semiconductors, provide platforms in which a discrete exciton couples to a confined electromagnetic mode and hybridizes into exciton-polaritons \cite{weisbuch1992observation,liu2015strong,dufferwiel2015exciton,flatten2016room,lidzey1998strong,kena2013ultrastrongly}. When the exciton--photon coupling \(g\) exceeds the relevant radiative and nonradiative decay rates, the system enters the strong-coupling regime, in which light and matter exchange energy coherently before decaying. This enables coherent control of quantum states and presents potential applications in modern quantum information science and technology \cite{nielsen2010quantum,hanson2008coherent,volz2012ultrafast,stute2013quantum,degen2017quantum}.

Several excitonic platforms have reached the ultrastrong-coupling regime, where \(g\) is a significant fraction of the bare cavity or exciton angular frequency, commonly quantified by \(\eta \equiv g/\max\{\omega_c,\omega_x\} \gtrsim 0.1\), with \(\omega_c\) and \(\omega_x\) denoting the bare cavity and exciton angular frequencies, respectively. In this regime the rotating-wave approximation fails, counter-rotating processes become relevant, and the ground state is a dressed vacuum with nonzero virtual photon and exciton populations \cite{ciuti2005quantum,frisk2019ultrastrong,forn2019ultrastrong}.

The exciton--photon coupling in microcavities is set by the electromagnetic-field amplitude and the oscillator-strength density of the active medium. Stronger coupling is obtained by minimizing the effective mode volume and increasing the number of resonant dipoles within the optical mode. Advances in nanofabrication now provide slot resonators, photonic-crystal and metasurface cavities, and plasmonic gap structures that confine the field at the nanoscale and enhance the electromagnetic field \cite{almeida2004guiding,akahane2003high,zhang2018photonic,chen2020metasurface,weber2023intrinsic,chikkaraddy2016single,torma2014strong}. Further developments in microcavity design provide ultrashort cavity lengths with control of detuning and mode volume \cite{dufferwiel2014strong,dufferwiel2015exciton}. Collectively, these advances establish exciton--photon cavities as robust ultrastrong-coupling platforms operating at or near ambient conditions.

The ultrastrong-coupling regime in excitonic systems has already been realized experimentally across multiple material platforms.
Organic microcavities hosting Frenkel excitons achieve giant vacuum Rabi splittings at room temperature \cite{kena2013ultrastrongly,chang2023ultrastrong}. Device implementations show narrow-angle, stable emission governed by polaritonic dispersion engineering \cite{mischok2024breaking,mischok2023highly}. Metasurface-assisted and plasmonic resonators in van der Waals semiconductors realize ultrastrong coupling at room temperature with tunable detuning \cite{wu2024ultrastrong}. Layered halide perovskite microcavities leverage large oscillator strength to modify exciton kinetics, including the suppression of annihilation channels \cite{fei2024controlling}.

Importantly, these established experimental demonstrations show that the ultrastrong-coupling regime addressed in this work is not a purely mathematical model, but a physically relevant and experimentally accessible operating regime. Therefore, studying Kerr-induced optical bistability in ultrastrong coupling provides experimentally testable predictions for current microcavity and nanophotonic architectures, rather than an abstract extension of the conventional strong-coupling theory.

Excitonic systems exhibit an optical Kerr nonlinearity arising from exciton--exciton interactions. In an effective Hamiltonian description this appears as a quartic term. In a driven-dissipative cavity within the strong-coupling regime, it produces optical bistability with S-shaped input-output characteristics and hysteresis \cite{baas2004optical}, making the system not only state-dependent but also history-dependent. This bistability has served both as a probe of interactions and as a basis for low-power optical switching and memory \cite{lecomte2014polariton,leyder2007interference,cerna2013ultrafast,ballarini2013all}.  Related theoretical work has also explored optical bistability, multistability, and controllable nonlinear effects in coupled quantum-well and optomechanical resonators \cite{jabri2023optical,sete2012controllable}.
In the ultrastrong-coupling regime, the breakdown of the rotating-wave approximation enables counter-rotating processes that dress the ground state and can generate virtual excitations. These features make the system response more sensitive to the light--matter coupling strength, motivating us to examine whether and how they affect the onset of optical bistability and the corresponding bistability condition. To address this question over a broad range of coupling strengths, it is essential to start from a ``full'' ultrastrong model that retains counter-rotating light--matter processes (and the associated $A^{2}$ term) beyond the RWA approximation.

In this work we analyze the stability of a driven exciton--photon microcavity with a Kerr interaction under coherent optical pumping and dissipation in the ultrastrong-coupling regime. We determine the steady states, analyze their linear stability, and map the resulting stability regions. Our goal is to establish how the bistability of the cavity-exciton system changes in the ultrastrong-coupling regime relative to the strong-coupling regime. This framework allows us to track how the turning points and hysteresis window evolve as the coupling $g$ is varied, and to connect smoothly to the conventional strong-coupling prediction in the small \(\eta\) limit.

\section{Ultrastrong cavity-exciton model}
We consider a single confined cavity mode \(\hat a\) of angular frequency \(\omega_c\) coupled to a single bosonic exciton \(\hat b\) of angular frequency \(\omega_x\). The operators satisfy the usual bosonic commutation relations \([\hat a,\hat a^\dagger]=[\hat b,\hat b^\dagger]=1\). We set \(\hbar=1\).

We note that real excitons are composite electron--hole excitations, so treating \(\hat b\) as a perfect,
structureless boson is an effective approximation valid in the dilute regime \cite{combescot2008many,carusotto2013quantum,porras2002polariton}. Equivalently, this dilute-bosonic approximation requires the mean inter-exciton spacing to remain much larger than the exciton Bohr radius \(a_B\); once the spacing becomes comparable to \(a_B\), phase-space-filling effects invalidate the picture of structureless bosons. In many microcavity platforms,
an important source of nonlinearity is phase-space filling associated with the fermionic
nature of the underlying electron and hole states, which reduces the excitonic oscillator strength and therefore renormalizes the effective
light--matter coupling with increasing exciton density 
\cite{houdre1995saturation,rossbach2013impact,yagafarov2020mechanisms,porras2002polariton,richard2026excitonic}.
In the present work we focus on the dilute/weakly nonlinear regime and approximate the coupling by a constant \(g\), while capturing the leading nonlinear renormalization of the driven response via an effective Kerr-type interaction; incorporating an explicit coupling saturation \(g(n_x)\) is beyond the scope of the present model.

In the ultrastrong-coupling regime, the rotating-wave approximation fails; we therefore use the quantum Hopfield--Rabi model, which retains both rotating and counter-rotating exciton--photon processes. We write the total Hamiltonian as
\begin{equation}
\hat H = \hat H_{\mathrm{HR}} + \hat H_{\mathrm{Kerr}} + \hat H_{\mathrm{drive}} .
\end{equation}
This Hamiltonian is a sum of three parts: (i) \(\hat H_{\mathrm{HR}}\) describes a single exciton coupled to the cavity with strength \(g\), (ii) \(\hat H_{\mathrm{Kerr}}\) captures exciton--exciton interactions through the Kerr nonlinearity, and (iii) \(\hat H_{\mathrm{drive}}\) represents coherent driving. The Hopfield--Rabi Hamiltonian is:
\begin{equation}
\hat H_{\mathrm{HR}}
= \omega_c\,\hat a^\dagger \hat a
+ \omega_x\,\hat b^\dagger \hat b
+ g\,(\hat a+\hat a^\dagger)(\hat b+\hat b^\dagger)
+ D\,(\hat a+\hat a^\dagger)^2 .
\end{equation}

The contribution \(D(\hat a+\hat a^\dagger)^2\) arises from minimal coupling \cite{forn2019ultrastrong}. For the single bosonic matter mode of frequency \(\omega_x\) considered here, the Thomas--Reiche--Kuhn (TRK) sum rule \cite{thomas1925naturwissenschaften,kuhn1925gesamtstarke,reiche1925zahl} constrains the coefficient to \(D = g^2/\omega_x\) \cite{frisk2019ultrastrong,tufarelli2015signatures,larson2021jaynes}. Thus, although \(D(\hat a+\hat a^\dagger)^2\) is purely photonic in operator form, its prefactor is not purely photonic: it is fixed by the matter oscillator strength entering the same light--matter interaction that determines \(g\). Including this term is crucial in the ultrastrong-coupling regime, where it prevents the unphysical superradiant phase transition predicted by Dicke-type models when the \(A^2\) contribution is neglected \cite{wang1973phase,rzazewski1975phase,bialynicki1979no}.

In the ultrastrong-coupling regime, the excitonic Kerr interaction must be treated
in its full quadrature form, retaining all bosonic terms, as demonstrated by
Mauceri \emph{et al.}~\cite{mauceri2022ultrastrong}; in their work, the excitonic Kerr
nonlinearity is obtained from the leading quartic anharmonicity, while higher-order nonlinearities
are neglected as subleading in the weakly nonlinear regime. We follow the same modelling here.
Accordingly, we write the Kerr interaction explicitly as
\begin{equation}
\hat H_{\mathrm{Kerr}}
= \frac{J_b\omega_c}{6}\,(\hat b+\hat b^\dagger)^{4},
\label{eq:HKerrBare}
\end{equation}
where \(J_b\) quantifies the Kerr interaction strength. The prefactor \(1/6\) is chosen so that the number-conserving term
\(\hat b^{\dagger 2}\hat b^{2}\) appears with coefficient \(J_b\omega_c\) in the weak-coupling/RWA limit.

In the strong-coupling regime it is common to apply a rotating-wave approximation (RWA) to the quartic interaction, keeping only the number-conserving Kerr term \(\propto \hat b^{\dagger 2}\hat b^2\). In the ultrastrong-coupling regime, however, this truncation is not valid: counter-rotating contributions are not perturbatively small, so the bare excitation number is no longer conserved. We therefore retain the full quadrature form in Eq.~\eqref{eq:HKerrBare}, consistent with the microscopic derivation of Ref.~\cite{mauceri2022ultrastrong}. In the weak-coupling limit, applying the RWA to Eq.~\eqref{eq:HKerrBare} recovers the standard Kerr Hamiltonian (up to frequency-renormalization terms and an additive constant).

The driving field is given by
\begin{equation}
\hat H_{\mathrm{drive}}
= F_0\left(
  \hat a\, e^{-i \omega_d t}
  + \hat a^\dagger e^{i \omega_d t}
\right),
\end{equation}
with amplitude \(F_0\) and drive frequency \(\omega_d\).

To analyze the dynamics, we diagonalize \(\hat H_{\mathrm{HR}}\) via a Hopfield--Bogoliubov transformation (the derivation as well as the analytical expressions for the coefficients are given in Appendix~A), introducing bosonic polaritons \(\hat P_n\) (\(n=1,2\)), often referred to as the lower and upper polaritons \cite{basov2020polariton}.
\begin{equation}
\hat P_n = A_n^{*}\hat a + B_n^{*}\hat b - A_n'\hat a^\dagger - B_n'\hat b^\dagger,
\qquad
[\hat P_m,\hat P_n^\dagger] = \delta_{mn},
\label{eq:HopfieldDef}
\end{equation}
with inverse relations
\begin{equation}
\hat a = \sum_{n=1,2}\!\big(A_n \hat P_n + A_n' \hat P_n^\dagger\big),\qquad
\hat b = \sum_{n=1,2}\!\big(B_n \hat P_n + B_n' \hat P_n^\dagger\big),
\label{eq:InverseHopfield}
\end{equation}
we obtain the diagonal quadratic Hamiltonian
\begin{equation}
\hat H_{\mathrm R} = \sum_{n=1,2}\omega_n\,\hat P_n^\dagger \hat P_n,
\label{eq:HRdiag}
\end{equation}
where the polariton frequencies are ordered as \(\omega_1 \le \omega_2\).

The normal‑mode frequencies satisfy
\begin{equation}\label{eq:polfreq2}
\omega_{1,2}^{\,2}
= \tfrac12\!\left(\omega_c^2+4D\omega_c+\omega_x^2
\mp \sqrt{\bigl(\omega_c^2+4D\omega_c-\omega_x^2\bigr)^2
          +16g^2\,\omega_c\omega_x}\right)\,.
\end{equation}

\begin{figure}[t]
    \centering
    \includegraphics[width=\linewidth]{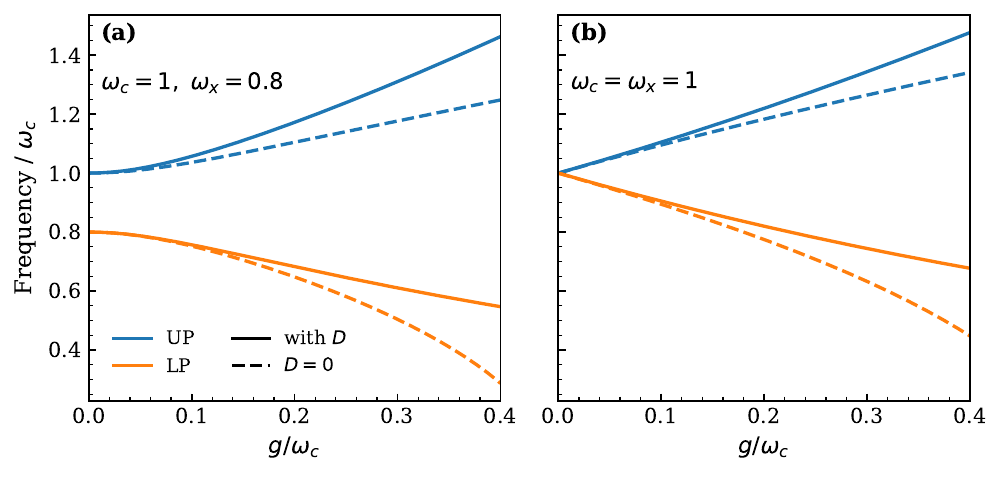}
    \caption{Upper (UP) and lower (LP) polariton eigenfrequencies
    \(\omega_{1,2}\) (normalized to \(\omega_c\)) as a function of the normalized
    light--matter coupling strength \(g/\omega_c\).
    (a) Detuned case \((\omega_c = 1,\ \omega_x = 0.8)\).
    (b) Resonant case \((\omega_c = \omega_x = 1)\).
    In both panels, solid lines correspond to the full model including the
    diamagnetic term \(D\), while dashed lines correspond to the case where the diamagnetic term is neglected (\(D=0\)).
    Blue and orange curves denote the UP and LP branches, respectively.
    Including \(D\) produces an overall blueshift of both polariton branches,
    which becomes more pronounced at larger coupling.}    \label{fig:polariton_frequencies}
\end{figure}

The dependence of the polariton frequencies on the light--matter coupling
strength \(g\) and the diamagnetic coupling \(D\) is explicit in
Eq.~\eqref{eq:polfreq2}. For fixed bare frequencies \(\omega_c\) and
\(\omega_x\), increasing \(g\) enhances the cavity--exciton hybridization and
widens the splitting between \(\omega_1\) and \(\omega_2\). The diamagnetic term
\(D\) renormalizes the bare cavity frequency,
\(\omega_c \to \tilde{\omega}_c=\sqrt{\omega_c^2+4D\omega_c}\), leading 
to a blueshift of both branches. This is illustrated in
Fig.~\ref{fig:polariton_frequencies}(a,b) by comparing spectra with \(D\) (solid)
and without \(D\) (dashed) for the detuned (a) and resonant (b) cases; in both,
the blueshift becomes more pronounced as \(g\) increases.

Rewriting Eq.~\eqref{eq:HKerrBare} in the polariton basis using Eq.~\eqref{eq:InverseHopfield}, we obtain
\begin{equation}
\hat H_{\mathrm{Kerr}}
= \frac{J_b\omega_c}{6}
\left(\sum_{n=1,2} C_n \hat P_n + C_n^* \hat P_n^\dagger\right)^{\!4},
\qquad
C_n \equiv B_n + B_n'^* ,
\label{eq:HKerrPolaritonFull}
\end{equation}
and the drive Hamiltonian becomes
\begin{equation}
\hat H_{\mathrm{drive}}
= \sum_{n=1,2} F_0\!\left[
\big(A_n e^{-i\omega_d t}+A_n'^* e^{i\omega_d t}\big)\hat P_n
+\big(A_n' e^{-i\omega_d t}+A_n^{*} e^{i\omega_d t}\big)\hat P_n^\dagger
\right].
\label{eq:HdrivePolaritonFull}
\end{equation}

\subsection*{Lower–polariton approximation}

We now focus on a parameter regime where the lower and upper polariton frequencies are
well separated. The single--lower--polariton
reduction is valid provided that the Hamiltonian scales capable of coupling the lower
and upper branches remain small compared with both the inter-branch splitting
\(|\omega_2-\omega_1|\) and the upper-polariton detuning from the drive
\(|\omega_2-\omega_d|\).
This occurs either in the strong-coupling regime, where large values of the
light--matter coupling \(g\) lead to a larger separation between the
polariton frequencies, or when the bare exciton and cavity are initially
strongly detuned, i.e.\ for large \(|\omega_x - \omega_c|\).
For example, using Eq.~\eqref{eq:A8}, when the bare modes are resonant
(\(\omega_x=\omega_c\)) the inter-branch splitting reduces to
\(|\omega_2-\omega_1|=2g\), so achieving \(|\omega_2-\omega_1|\simeq 0.1\,\omega_x\)
corresponds to \(g/\omega_c\simeq 0.05\). In contrast, for a detuned cavity a
comparable splitting can be obtained with a smaller \(g/\omega_c\), because the
bare detuning itself contributes to \(|\omega_2-\omega_1|\). In either case, the
upper branch can be neglected provided that the Kerr-induced nonlinear energy
scale remains much smaller than the inter-branch splitting, e.g.\
\(J_b\omega_c |C_1|^4 \langle \hat P_1^\dagger \hat P_1\rangle \ll |\omega_2-\omega_1|\),
and that the near-resonant drive matrix element for the upper polariton,
\(F_0|A_2'|\), remains much smaller than its detuning from the drive,
\(|\omega_2-\omega_d|\),
so that Kerr-mediated scattering processes that would populate the upper
polariton are far off-resonant.

In this regime, the upper polariton lies far outside the relevant dynamical range and can therefore be neglected.
Moreover, the virtual counter-rotating effects of the ultrastrong-coupling
quadratic Hamiltonian are not discarded by this step: they are already encoded in the
renormalized polariton frequencies \(\omega_{1,2}\) and Hopfield coefficients.
This well-separated-branch regime is experimentally feasible, for example by operating in a detuned cavity--exciton system, which suppresses inter-branch scattering processes.

Keeping only the lower polariton, we write
\begin{equation}
    \hat P_1
    = A_1^{*}\hat a + B_1^{*}\hat b
      - A_1' \hat a^\dagger - B_1' \hat b^\dagger,
    \qquad
    [\hat P_1,\hat P_1^\dagger] = 1,
\end{equation}
with the inverse relations 
\begin{equation}
    \hat a = A_1 \hat P_1 + A_1' \hat P_1^\dagger,
    \qquad
    \hat b = B_1 \hat P_1 + B_1' \hat P_1^\dagger .
\end{equation}
In this lower–polariton approximation, the quadratic Hopfield–Rabi
Hamiltonian becomes
\begin{equation}
    \hat H_{\mathrm R}
    = \omega_1\,\hat P_1^\dagger \hat P_1.
\end{equation}
The Kerr interaction reduces to
\begin{equation}
    \hat H_{\mathrm{Kerr}}
    = \frac{J_b\,\omega_c}{6}
      \left(C_1 \hat P_1 + C_1^* \hat P_1^\dagger\right)^{4},
    \qquad
    C_1 \equiv B_1 + B_1'^* ,
\end{equation}
and the drive becomes
\begin{equation}
    \hat H_{\mathrm{drive}}
    = F_0\!\left[
      \big(A_1 e^{-i\omega_d t}+A_1'^* e^{i\omega_d t}\big)\hat P_1
      +\big(A_1' e^{-i\omega_d t}+A_1^{*} e^{i\omega_d t}\big)\hat P_1^\dagger
      \right].
\end{equation}
Collecting all contributions, the system in the regime
\(\omega_1 \ll \omega_2\) is described by the effective single–mode
lower–polariton Hamiltonian
\begin{equation}
    \hat H_{\mathrm{LP}}
    \simeq
    \omega_1 \hat P_1^\dagger \hat P_1
    + \frac{J_b\,\omega_c}{6}
      \left(C_1 \hat P_1 + \text{H.c.}\right)^{4}
    + F_0\!\left[
      \big(A_1 e^{-i\omega_d t}+A_1'^* e^{i\omega_d t}\big)\hat P_1
      + \text{H.c.}
    \right].
\end{equation}
In the rotating frame of the driving field at frequency \(\omega_d\), defined by
\(\hat U(t) = \exp\!\big(i \omega_d t\,\hat P_1^\dagger \hat P_1\big)\), and neglecting
the fast-oscillating terms in the polariton basis, we obtain the effective
lower--polariton Hamiltonian
\begin{equation}
    \hat H_{\mathrm{eff}}
    =
    \tilde{\Delta}_1\,\hat P_1^\dagger \hat P_1
    + U_{\mathrm{LP}}\,\hat P_1^{\dagger 2}\hat P_1^{2}
    + \left(\mathcal{F}_1 \hat P_1 + \mathcal{F}_1^* \hat P_1^\dagger\right),
\end{equation}
where
\begin{equation}
    \tilde{\Delta}_1 = \omega_1 - \omega_d + 2U_{\mathrm{LP}},\qquad
    U_{\mathrm{LP}} =  J_b\,\omega_c\,|C_1|^4,\qquad
    \mathcal{F}_1 = F_0 A_1'^* .
\end{equation}
Here the rotating-frame averaging is performed in the polariton basis.
A sufficient condition for the rotating-wave step to be valid is that
the slow dynamical scales of the driven Kerr problem remain small compared to the harmonic
frequencies \(2\omega_d\) and \(4\omega_d\), i.e.
\begin{equation}
\max\!\left\{
|F_{0}A_{1}|,\;
U_{\mathrm{LP}}\,n_{1}
\right\}
\ll 2\omega_d .
\label{eq:RWA_condition}
\end{equation}
where \(n_{1}\equiv \langle \hat P_1^\dagger \hat P_1\rangle\) denotes the lower--polariton
occupation number. For sufficiently strong driving such that
Eq.~\eqref{eq:RWA_condition} is not satisfied, one must retain the full time-dependent Hamiltonian.

\section{Stability Analysis}

To investigate the dynamical response and bistability of the driven lower polariton, we now include losses and treat the system as an open quantum system. The system density matrix \(\hat\rho\) obeys a Lindblad master equation
\begin{equation}
    \dot{\hat\rho}
    = -i[\hat H_{\mathrm{eff}},\hat\rho]
      + \kappa_1(g)\,\mathcal{D}[\hat P_1]\hat\rho,
\end{equation}
where \(\kappa_1(g)\) denotes the lower--polariton decay rate and
\begin{equation}
    \mathcal{D}[\hat O]\hat\rho
    = \hat O \hat\rho \hat O^\dagger - \tfrac12\{\hat O^\dagger \hat O,\hat\rho\}
\end{equation}
is the usual Lindblad dissipator.

In the ultrastrong-coupling regime, where varying \(g\) can shift the polariton resonances over a wide frequency range, it is important to account for the frequency dependence of the dissipation rate. For an infinitely thin metallic mirror, the Maxwell-boundary-condition (MBC) analysis leads to the dissipation rate~\cite{bamba2013system,de2014comment}:
\begin{equation}
\kappa_1(g)
\simeq \frac{\kappa_{0}}{1+\left[\omega_x/\omega_1(g)\right]^2}.
\label{eq:kappa1g}
\end{equation}

where \(\kappa_{0}\) is a \(g\)-independent dissipation-rate set by mirror losses.

We use a Markovian Lindblad description because the damping is weak, \(\kappa_1(g)\ll \omega_1(g)\), corresponding to coupling to a broadband photonic
continuum with negligible reservoir memory on the polariton lifetime scale.

In the following, we focus on the semiclassical regime and consider directly the coherent lower--polariton amplitude \(\langle \hat P_1\rangle(t)\). Within a mean-field approximation we factorize higher-order correlators according to
\begin{equation}
    \langle \hat P_1^{\dagger}\hat P_1\hat P_1\rangle \approx
    \langle \hat P_1^{\dagger}\rangle\langle \hat P_1\rangle^2
    = \big|\langle \hat P_1\rangle\big|^2 \langle \hat P_1\rangle,
\end{equation}
which yields the semiclassical equation of motion
\begin{equation}
    \frac{d}{dt}\langle \hat P_1\rangle
    = -\left(\frac{\kappa_1(g)}{2}+i\tilde{\Delta}_1\right)\langle \hat P_1\rangle
      - i\,2U_{\mathrm{LP}}\big|\langle \hat P_1\rangle\big|^2\langle \hat P_1\rangle
      - i\,\mathcal{F}_1^* .
    \label{eq:alpha_eom}
\end{equation}
Steady states \(\langle \hat P_1\rangle_{\mathrm{ss}}\) are defined by
\(d\langle \hat P_1\rangle/dt = 0\). Using Eq.~\eqref{eq:alpha_eom} we obtain
\begin{equation}
    \left(
      \frac{\kappa_1(g)}{2}
      + i\bigl[\tilde{\Delta}_1 + 2U_{\mathrm{LP}} \big|\langle \hat P_1\rangle_{\mathrm{ss}}\big|^2\bigr]
    \right)\langle \hat P_1\rangle_{\mathrm{ss}}
    = -i\,\mathcal{F}_1^* .
\end{equation}
Taking the modulus squared of both sides leads to
\begin{equation}
    |\mathcal{F}_1|^2
    = \big|\langle \hat P_1\rangle_{\mathrm{ss}}\big|^2
      \left\{
        \bigl[\tilde{\Delta}_1 + 2U_{\mathrm{LP}} \big|\langle \hat P_1\rangle_{\mathrm{ss}}\big|^2\bigr]^2
        + \left(\frac{\kappa_1(g)}{2}\right)^2
      \right\}.
    \label{eq:input_output_alpha}
\end{equation}
Identifying the mean lower--polariton population with the squared modulus of the coherent amplitude, \(n \equiv \langle \hat P_1^\dagger \hat P_1\rangle \approx \big|\langle \hat P_1\rangle_{\mathrm{ss}}\big|^2\), Eq.~\eqref{eq:input_output_alpha} can be written as a nonlinear input--output relation between the steady-state polariton population \(n\) (output) and the drive intensity \(|\mathcal{F}_1|^2\) (input),
\begin{equation}
    I_{\mathrm{in}}(n) \equiv |\mathcal{F}_1|^2
    = n\left\{
      \bigl[\tilde{\Delta}_1 + 2U_{\mathrm{LP}} n\bigr]^2
      + \left(\frac{\kappa_1(g)}{2}\right)^2
    \right\},
    \label{eq:cubic_n}
\end{equation}
where \(I_{\mathrm{in}}(n)\) denotes the input intensity required to obtain
a steady-state lower--polariton population \(n\).

Optical bistability appears when \(I_{\mathrm{in}}(n)\) becomes nonmonotonic, such that the same input intensity corresponds to multiple steady-state populations. For a smooth nonlinear function, this is equivalent to the existence of turning points where the derivative is zero. The turning points of the input--output curve are therefore determined by the condition \(d I_{\mathrm{in}}/dn = 0\). Using Eq.~\eqref{eq:cubic_n}, this yields
\begin{equation}
    12 U_{\mathrm{LP}}^2 n^2
    + 8 \tilde{\Delta}_1 U_{\mathrm{LP}} n
    + \left(\tilde{\Delta}_1^{\,2} + \frac{\bigl(\kappa_1(g)\bigr)^{2}}{4}\right)
    = 0,
\end{equation}
with solutions
\begin{equation}
    n_{c,\pm}
    = \frac{-4\tilde{\Delta}_1
            \pm \sqrt{4\tilde{\Delta}_1^{\,2}-3\bigl(\kappa_1(g)\bigr)^{2}}}
           {12 U_{\mathrm{LP}}}.
    \label{eq:ncpm}
\end{equation}
Real turning points exist only if the discriminant is positive,
\begin{equation}
    4\tilde{\Delta}_1^{\,2}-3\bigl(\kappa_1(g)\bigr)^{2} > 0
    \quad\Rightarrow\quad
    |\tilde{\Delta}_1|
    > \frac{\sqrt{3}}{2}\,\kappa_1(g).
\end{equation}
Since the intensity is positive, and therefore \(n_{c,\pm}>0\), it is required that \(\tilde{\Delta}_1 < 0\).
Thus
\begin{equation}
    \omega_1 - \omega_d + 2U_{\mathrm{LP}} < 0.
\end{equation}

To determine which of these semiclassical steady states are physically realized, we now analyse their dynamical stability. We consider small fluctuations of the lower--polariton amplitude around a given steady state,
\begin{equation}
    \langle \hat P_1\rangle(t)
    = \langle \hat P_1\rangle_{\mathrm{ss}} + \delta P_1(t).
\end{equation}
Inserting this into Eq.~\eqref{eq:alpha_eom} and keeping only terms linear in
\(\delta P_1\) and \(\delta P_1^*\) yields
\begin{equation}
    \frac{d}{dt}\,\delta P_1
    = -\left(\frac{\kappa_1(g)}{2}
              + i\tilde{\Delta}_1\right)\delta P_1
      - i\,4U_{\mathrm{LP}} n\,\delta P_1
      - i\,2U_{\mathrm{LP}}\langle \hat P_1\rangle_{\mathrm{ss}}^{\,2}\,\delta P_1^* .
\end{equation}
Together with the complex conjugate equation for \(\delta P_1^*\), this can be written in matrix form as
\begin{equation}
    \frac{d}{dt}
    \begin{pmatrix}
        \delta P_1 \\
        \delta P_1^*
    \end{pmatrix}
    =
    \mathbf{M}
    \begin{pmatrix}
        \delta P_1 \\
        \delta P_1^*
    \end{pmatrix},
\end{equation}
where
\begin{equation}
    \mathbf{M}
    =
    \begin{pmatrix}
        -\dfrac{\kappa_1(g)}{2}
        - i\bigl(\tilde{\Delta}_1 + 4U_{\mathrm{LP}}n\bigr)
      & - i\,2U_{\mathrm{LP}}\langle \hat P_1\rangle_{\mathrm{ss}}^{\,2}
        \\[4pt]
        i\,2U_{\mathrm{LP}}\langle \hat P_1\rangle_{\mathrm{ss}}^{*\,2}
      & -\dfrac{\kappa_1(g)}{2}
        + i\bigl(\tilde{\Delta}_1 + 4U_{\mathrm{LP}}n\bigr)
    \end{pmatrix}.
    \label{eq:Mmatrix}
\end{equation}
The eigenvalues \(\lambda\) of \(\mathbf{M}\) determine the linear stability of
the steady state, it is stable if and only if both eigenvalues have negative
real parts. These eigenvalues satisfy the quadratic characteristic equation
\begin{equation}
    \lambda^2
    + \kappa_1(g) \lambda
    + a_0(n)
    = 0,
    \label{eq:charpoly_stability}
\end{equation}
with
\begin{equation}
    a_0(n)
    =
    12 U_{\mathrm{LP}}^{2} n^{2}
    + 8 \tilde{\Delta}_1 U_{\mathrm{LP}} n
    + \tilde{\Delta}_1^{\,2}
    + \frac{\bigl(\kappa_1(g)\bigr)^{2}}{4}.
\end{equation}

Since \(\kappa_1(g) > 0\) by definition of the decay rate, both eigenvalues have negative real parts iff \(a_0(n) > 0\). The function \(a_0(n)\) is a quadratic polynomial in \(n\), whose two real roots
are precisely the critical populations \(n_{c,\pm}\) defined in
Eq.~\eqref{eq:ncpm}. We can therefore factorize it as
\begin{equation}
    a_0(n)
    = 12 U_{\mathrm{LP}}^{2}\,(n - n_{c,-})(n - n_{c,+}) .
\end{equation}
Since the prefactor \(12 U_{\mathrm{LP}}^{2} > 0\), it follows that
\(a_0(n) > 0\) for \(n < n_{c,-}\) and \(n > n_{c,+}\), while \(a_0(n) < 0\) for
\(n_{c,-} < n < n_{c,+}\). Thus, in the bistable regime the lower and upper
branches of the S-shaped input--output curve are dynamically stable, whereas the intermediate
branch between \(n_{c,-}\) and \(n_{c,+}\) is unstable.

\begin{figure}
    \centering
    \includegraphics[width=\linewidth]{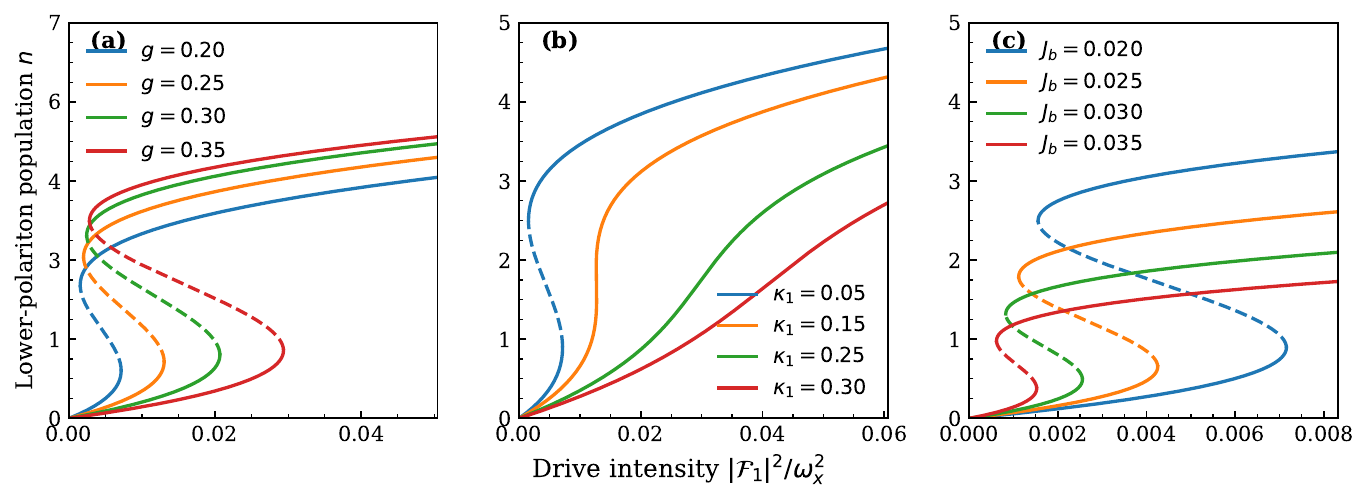}
    \caption{
    Input--output curves of the driven lower polariton, obtained from the
    nonlinear relation in Eq.~\eqref{eq:cubic_n}, with fixed bare
    frequencies \(\omega_c = 1.4\), \(\omega_x = 1.0\) and drive frequency
    \(\omega_d = 1.1\).
    The vertical axis shows the lower--polariton population \(n\), while
    the horizontal axis displays the drive intensity \(|\mathcal{F}_1|^2\).
    Solid lines indicate dynamically stable steady states, while dashed
    segments correspond to the unstable branch between the turning points.
    (a) Dependence on the light--matter coupling strength \(g\) at fixed
    \(\kappa_0 = 0.1\)
    and Kerr nonlinearity \(J_b = 0.02\); the different
    values of \(g\) are indicated in the legend.
    Increasing \(g\) lowers the lower--polariton frequency, makes the
    effective detuning more negative, and shifts the bistable S-shaped
    response to higher input intensities.
    (b) Dependence on the lower--polariton decay rate \(\kappa_{1}(g)\) at fixed
    \(g = 0.20\) and \(J_b = 0.02\).
    Larger \(\kappa_{1}(g)\) reduces the extent of the bistable region and eventually
    removes the turning points, leaving a single--valued response.
    (c) Dependence on the excitonic Kerr nonlinearity \(J_b\) at fixed
    \(g = 0.20\) and fixed \(\kappa_0 = 0.1\) (thus fixed \(\kappa_{1}(g)\)).
    Increasing \(J_b\) enhances the effective lower--polariton interaction
    \(U_{\mathrm{LP}}\), so that bistability appears at lower drive
    intensities, as reflected by the leftward shift of the S-shaped region.
    }
    \label{fig:figure_2}
\end{figure}

Figure~\ref{fig:figure_2} summarizes the input--output
response obtained from Eq.~\eqref{eq:cubic_n}. In the following, to provide a quantitative visual comparison in
dimensionless units, we evaluate Eq.~\eqref{eq:cubic_n} after normalizing all frequencies and rates
to the exciton frequency \(\omega_x\) (i.e., we set \(\omega_x=1\)), so that \(\omega_c\), \(\omega_d\),
\(g\), \(\kappa_1(g)\) and \(U_{\mathrm{LP}}\) are expressed in units of \(\omega_x\).
Throughout, we fix the bare
frequencies to \(\omega_c = 1.4\), \(\omega_x = 1.0\), and the drive to
\(\omega_d = 1.1\), so that all curves correspond to the same driving
frequency while the remaining parameters are varied.

With the parameters used in Fig.~\ref{fig:figure_2}
($\omega_c=1.4$, $\omega_x=1$, $\omega_d=1.1$, $\kappa_0=0.10$, $J_b=0.02$),
both approximations introduced above are well satisfied throughout the plotted range.

For the single--lower--polariton reduction, the drive is detuned from the upper
polariton by $\omega_2-\omega_d \simeq 0.34$--$0.56$. For the drive range used in
Fig.~\ref{fig:figure_2}, we find $n\lesssim 4.5$ and
$F_0|A_2'|=\left|\frac{A_2'}{A_1'}\right|\,|\mathcal F_1|$. Together with
Eq.~\eqref{eq:kappa1g}, this means that the linewidth, the omitted upper-polariton
drive scale, and the Kerr shift all remain small compared with $\omega_2$:
\begin{equation}
\max\!\left\{
\frac{\kappa_1(g)}{\omega_2},\;
\frac{F_0|A_2'|}{\omega_2},\;
\frac{2U_{\mathrm{LP}}\,n}{\omega_2}
\right\}
\lesssim 1.6\times10^{-1}.
\label{eq:UP_scales_small}
\end{equation}
In addition, the omitted upper-polariton drive term remains perturbative because
\(F_0|A_2'| \ll |\omega_2-\omega_d|\) throughout the same range.
Moreover, the condition used above,
$U_{\mathrm{LP}}n \ll |\omega_2-\omega_1|$, is also satisfied in the same range:
\begin{equation}
\frac{U_{\mathrm{LP}}\,n}{|\omega_2-\omega_1|}
\lesssim 1.6\times10^{-1}.
\label{eq:UP_mixing_small}
\end{equation}
This confirms that drive- and Kerr-mediated processes that would populate the upper polariton
remain far off resonance for Fig.~\ref{fig:figure_2}.

The polariton-basis rotating-wave step is likewise valid. In the rotating frame
at $\omega_d$, the terms neglected in the polariton basis oscillate at $2\omega_d$ and
$4\omega_d$, so we use Eq.~\eqref{eq:RWA_condition}. Using
$\mathcal F_1=F_0A_1'^*$ together with Eq.~\eqref{eq:A16}, the relevant drive scale can be written as
\begin{equation}
|F_0A_1|
=\Big|\frac{A_1}{A_1'}\Big|\,|\mathcal F_1|
=\frac{\omega_c+\omega_1}{\omega_c-\omega_1}\,|\mathcal F_1|
\lesssim 5.8\times10^{-1},
\qquad
\frac{\max\{|F_0A_1|,\;U_{\mathrm{LP}}n\}}{2\omega_d}\lesssim 2.6\times10^{-1}.
\label{eq:RWA_check_fig2}
\end{equation}
This supports the rotating-wave step used to obtain $\hat H_{\mathrm{eff}}$ in the regime shown.

In panel~(a) we tune
the light--matter coupling strength \(g\) at fixed bare cavity dissipation rate \(\kappa_0=0.10\) (so that \(\kappa_1(g)\) is given by Eq.~\eqref{eq:kappa1g})
and nonlinearity \(J_b\) (values indicated in the legend). As \(g\) increases,
the lower--polariton frequency \(\omega_1(g)\) decreases according to
Eq.~\eqref{eq:polfreq2}, and the effective detuning
\(\tilde{\Delta}_1 = \omega_1 - \omega_d + 2U_{\mathrm{LP}}\) becomes more
negative. This drives the system deeper into the bistable regime
\(|\tilde{\Delta}_1| > (\sqrt{3}/2)\,\kappa_1(g)\), shifts the critical populations
\(n_{c,\pm}\) of Eq.~\eqref{eq:ncpm} to larger values of
\(|\mathcal{F}_1|^2\), and broadens the S-shaped region of the
input--output curve.

Panels~(b) and (c) illustrate the competing roles of dissipation and Kerr
nonlinearity. In panel~(b) we vary the lower--polariton decay rate
\(\kappa_1(g)\) at fixed \(g\), \(J_b\), and \(\omega_d\)
(equivalently, we vary the bare cavity dissipation rate \(\kappa_0\) at fixed \(g\)).
Increasing
\(\kappa_1(g)\) suppresses the nonlinearity in Eq.~\eqref{eq:cubic_n} by
reducing the discriminant \(4\tilde{\Delta}_1^{\,2} - 3\bigl(\kappa_1(g)\bigr)^{2}\),
and thus shrinks the range of coupling over which turning points
exist. For sufficiently large \(\kappa_1(g)\), the two critical points merge
and the S-shaped response collapses to a single--valued curve. In
panel~(c) we instead vary the Kerr nonlinearity \(J_b\) at fixed \(g\),
fixed \(\kappa_0=0.10\), and \(\omega_d\). Since
\(U_{\mathrm{LP}} = J_b \omega_c |C_1|^{4}\), increasing \(J_b\)
enhances the effective lower--polariton interaction, lowers the critical
populations \(n_{c,\pm}\), and shifts the onset of bistability to
smaller drive intensities. Intuitively, stronger Kerr interactions imply
that fewer polaritons are required for the nonlinear shift
\(2U_{\mathrm{LP}}n\) to compete with the detuning \(\tilde{\Delta}_1\),
in agreement with the leftward motion of the S-shaped region in
panel~(c).

\begin{figure}
    \centering
    \includegraphics[width=\linewidth]{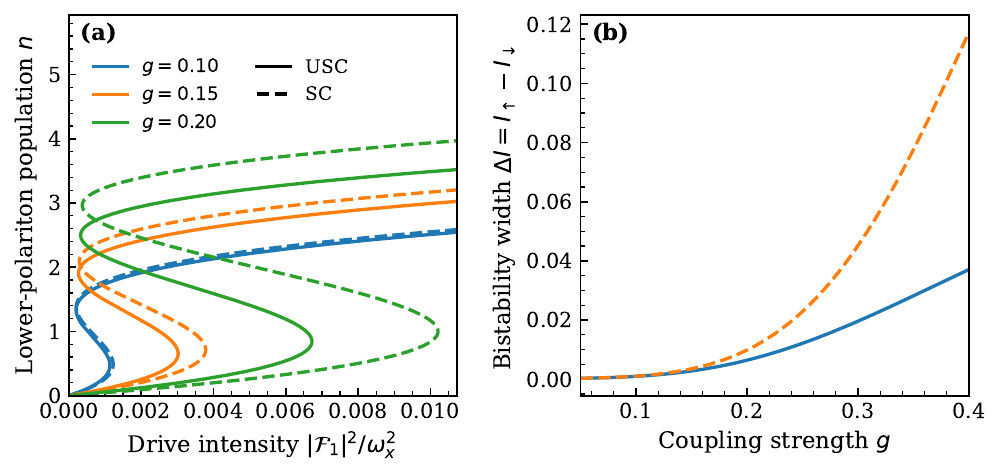}
    \caption{
    Comparison between the full ultrastrong-coupling (USC) response and the
    strong-coupling (SC) approximation \cite{baas2004optical} for the driven
    lower polariton, obtained from the nonlinear relation in
    Eq.~\eqref{eq:cubic_n}, with fixed bare frequencies \(\omega_c = 1.4\),
    \(\omega_x = 1.0\), and drive frequency \(\omega_d = 1.1\).
    Throughout, we fix the Kerr nonlinearity to \(J_b = 0.02\) and the bare
    cavity dissipation rate to \(\kappa_0 = 0.05\).
    Solid (dashed) lines denote the USC (SC) result.
    (a) Input--output curves of the driven lower polariton, where the vertical
    axis shows the lower--polariton population \(n\) and the horizontal axis
    displays the drive intensity \(|\mathcal{F}_1|^2/\omega_x^2\); the
    different values of \(g\) are indicated in the legend.
    At small light--matter coupling, the two descriptions nearly coincide,
    whereas at larger \(g\) the SC approximation predicts a broader S-shaped
    response and a larger bistable region than the full model.
    (b) Corresponding bistability width
    \(\Delta I = I_{\uparrow}-I_{\downarrow}\) as a function of the coupling
    strength \(g\).
    The deviation remains small in the weak-coupling regime but grows rapidly
    at larger \(g\), showing that the full USC model is required to capture
    the complete evolution of bistability over the entire coupling range.
    }
    \label{fig:figure_3}
\end{figure}

To assess the range of validity of the strong-coupling (SC) approximation \cite{baas2004optical}, we compare in Fig.~\ref{fig:figure_3} the full ultrastrong-coupling (USC) response with its SC approximation. As shown in Fig.~\ref{fig:figure_3}(a), the two descriptions nearly coincide at small light--matter coupling, indicating that the SC treatment accurately captures the weak-coupling regime. As \(g\) increases, however, clear discrepancies emerge: the SC approximation predicts a broader bistable region and, consequently, a larger bistability width \(\Delta I = I_{\uparrow}-I_{\downarrow}\), where \(I_{\uparrow}\) and \(I_{\downarrow}\) denote the upper and lower switching intensities, respectively, than the full USC model.

This is quantified in Fig.~\ref{fig:figure_3}(b), where the deviation remains small at weak coupling but grows rapidly in the large-coupling regime. The main quantitative difference between the two approaches therefore appears at large \(g\), showing that the full USC model is required to capture the complete evolution of bistability over the entire coupling range. This behavior is also consistent with the expected range of validity of the SC approximation: since the SC description breaks down at large \(\eta\), one expects the bistable region to deviate from that predicted by the USC model.

\section{Conclusions}

We have presented a microscopic description of optical bistability for a coherently driven
exciton--polariton mode in the ultrastrong--coupling regime. Starting from the
Hopfield--Rabi Hamiltonian with a full-quadrature excitonic Kerr interaction
and working in a parameter regime where the polariton branches are well separated,
\(\omega_1 \ll \omega_2\), we derived an effective single--mode model in which the lower
polariton is described by a driven Kerr Hamiltonian with renormalized frequency
\(\omega_1\), nonlinearity \(U_{\mathrm{LP}}\), and drive amplitude
\(\mathcal{F}_1\). Within a semiclassical treatment of the corresponding
Lindblad master equation, we obtained a cubic input--output relation,
Eq.~\ref{eq:cubic_n}, from which we derived the analytical bistability
condition \(|\tilde{\Delta}_1| > (\sqrt{3}/2)\,\kappa_1\) and the critical
populations \(n_{c,\pm}\). Our analysis shows how the competition between
light--matter coupling, Kerr nonlinearity, and loss controls the existence and
shape of the S--shaped response, and clarifies that, in the ultrastrong regime,
bistability is most naturally understood in the polariton representation rather
than in the bare exciton--photon basis.
Although the semiclassical bistability criterion retains its standard Kerr--oscillator form, the ultrastrong-coupling renormalization changes the effective parameters $g$-dependent, shifting the turning points and hysteresis window relative to the strong-coupling (RWA) prediction.
Starting from the full ultrastrong-coupling model allows the framework to describe bistability continuously as the coupling strength is varied, while recovering the conventional strong-coupling limit in the small-\(\eta\) regime.

A key implication of the present study is that it allows the bistable response to be studied across a broad range of light--matter coupling strengths \(g\), from the conventional strong-coupling regime into the ultrastrong-coupling regime. In this way, one can track how device-relevant quantities such as the switching thresholds, turning points, and hysteresis width evolve as \(g\) is varied. The bistable region can be engineered via the light--matter
coupling strength \(g\). This tunability originates from the explicit \(g\)-dependence of both
the lower--polariton frequency and the effective Kerr interaction, which enter the
bistability condition through \(\tilde{\Delta}_1=\omega_1-\omega_d+2U_{\mathrm{LP}}\) and
\(U_{\mathrm{LP}}=J_b\omega_c|C_1|^4\). Since \(g\) is set by electromagnetic confinement and
oscillator-strength density through the mode volume and the number or density of resonant dipoles,
our results provide a direct device-level knob to set and optimize switching thresholds and hysteresis widths across this wide coupling range.
This is directly relevant for bistability-based applications such as all-optical switching and optical memory,
where the two stable branches represent low- and high-output states and the hysteresis loop provides a finite
noise margin. A larger hysteresis width, which can be achieved by increasing \(g\) to widen the bistable region,
is favorable for memory-related applications because it increases the separation between the set and reset thresholds
and improves the noise margin \cite{cerna2013ultrafast,nakarmi2014analysis}. A smaller hysteresis width is more suitable
for fast, reversible switching and transistor-like logic, where pronounced history dependence is undesirable
\cite{ballarini2013all,nakarmi2014analysis}.

We emphasize that these conclusions are derived within the dilute/weakly nonlinear regime, in which the light--matter coupling can be treated as a constant parameter.
Incorporating phase-space filling would make the coupling density dependent, and would lead to a corresponding density dependence of the polaritonic composition; this extension is left for future work.

It will be interesting to extend this work beyond the single lower--polariton approximation in regimes
where the Kerr-induced nonlinear energy scale is no longer negligible compared to the inter-branch
splitting \( |\omega_2-\omega_1| \). In that case, both polariton branches can participate in the
dynamics, and an effective nonlinear description should include not only the self-Kerr (intra-branch)
nonlinearities of each branch but also the cross-Kerr (inter-branch) nonlinearity coupling the two
populations, potentially leading to a richer stability landscape.

\section{Declaration of competing interest}
The authors state that they have no financial or personal conflicts of interest that could have influenced the work presented in this article.

\section{Availability of Data}
The data supporting the findings of this study are available from the corresponding author upon reasonable request.

\section{Acknowledgments}
This work was supported by Tamkeen through the NYU Abu Dhabi Research Institute grant CG008 and by the Kawader Fellowship 
Program.

\appendix

\section{Bogoliubov transformation}

\paragraph{Normal–mode frequencies.}
Introduce the canonical quadratures
\begin{equation}\label{eq:A1}
\begin{aligned}
\hat X_a &= \frac{\hat a+\hat a^\dagger}{\sqrt{2\omega_c}},
&\qquad
\hat\Pi_a &= -i\sqrt{\frac{\omega_c}{2}}(\hat a-\hat a^\dagger),\\
\hat X_b &= \frac{\hat b+\hat b^\dagger}{\sqrt{2\omega_x}},
&\qquad
\hat\Pi_b &= -i\sqrt{\frac{\omega_x}{2}}(\hat b-\hat b^\dagger),
\end{aligned}
\end{equation}
which satisfy the canonical commutation relations
\begin{equation}\label{eq:A2}
[\hat X_\mu,\hat\Pi_\nu]=i\,\delta_{\mu\nu}.
\end{equation}
In these operators, the quantum Rabi Hamiltonian becomes
\begin{equation}\label{eq:A3}
\hat H_{\mathrm {HR}}
=\tfrac{1}{2}(\hat\Pi_a^2+\omega_c^2\hat X_a^2)
+\tfrac{1}{2}(\hat\Pi_b^2+\omega_x^2\hat X_b^2)
+2g\sqrt{\omega_c\omega_x}\,\hat X_a\hat X_b
+2D\,\omega_c\,\hat X_a^{2}.
\end{equation}
Applying Hamilton’s equations of motion,
\begin{equation}\label{eq:A4}
\dot{\hat X}_\mu=\frac{\partial \hat H_{\mathrm R}}{\partial \hat\Pi_\mu},
\qquad
\dot{\hat\Pi}_\mu=-\frac{\partial \hat H_{\mathrm R}}{\partial \hat X_\mu},
\end{equation}
one obtains the first‑order coupled system
\begin{equation}\label{eq:A5}
\begin{aligned}
\dot{\hat X}_a&=\hat\Pi_a,
&\qquad
\dot{\hat\Pi}_a&=-\,(\omega_c^2+4D\omega_c)\,\hat X_a
                 -2g\sqrt{\omega_c\omega_x}\,\hat X_b,\\
\dot{\hat X}_b&=\hat\Pi_b,
&\qquad
\dot{\hat\Pi}_b&=-\,\omega_x^2 \hat X_b
                 -2g\sqrt{\omega_c\omega_x}\,\hat X_a.
\end{aligned}
\end{equation}
Taking a second time derivative yields the second‑order matrix oscillator equation
\begin{equation}\label{eq:A6}
\ddot{\mathbf{\hat X}}+K\,\mathbf{\hat X}=0,
\qquad
\mathbf{\hat X}=
\begin{pmatrix}
\hat X_a\\ \hat X_b
\end{pmatrix},
\quad
K=
\begin{pmatrix}
\omega_c^2+4D\omega_c & 2g\sqrt{\omega_c\omega_x}\\[3pt]
2g\sqrt{\omega_c\omega_x} & \omega_x^2
\end{pmatrix}.
\end{equation}
The normal‑mode ansatz \(\mathbf{\hat X}(t)\propto e^{-i\omega t}\) gives the eigenvalue equation
\begin{equation}\label{eq:A7}
\det(K-\omega^2 \mathbb{1}_2)=0
\;\Longleftrightarrow\;
(\omega_c^2+4D\omega_c-\omega^2)(\omega_x^2-\omega^2)-4g^2\omega_c\omega_x=0,
\end{equation}
and hence the polariton frequencies
\begin{equation}\label{eq:A8}
\omega_{1,2}^{\,2}
=\tfrac12\!\left(\omega_c^2+4D\omega_c+\omega_x^2
\mp \sqrt{\bigl(\omega_c^2+4D\omega_c-\omega_x^2\bigr)^2
          +16g^2\,\omega_c\omega_x}\right)\,.
\end{equation}

\paragraph{Bogoliubov coefficients.}
To relate the polariton operators to the original cavity and matter operators,
we first diagonalize the oscillator matrix \(K\) and introduce the normal–mode
quadratures \(\hat X_n,\hat\Pi_n\) (\(n=1,2\)).
In terms of these variables, the bare quadratures can be written as
\begin{equation}\label{eq:A9}
\hat X_\mu=\sum_{n=1,2}u_{\mu n}\,\hat X_n,
\qquad
\hat\Pi_\mu=\sum_{n=1,2}u_{\mu n}\,\hat\Pi_n,
\qquad
\mu=a,b,
\end{equation}
where \(\hat X_\mu,\hat\Pi_\mu\) are the quadratures in the bare (cavity–matter)
basis, while \(\hat X_n,\hat\Pi_n\) are the quadratures in the polariton basis.
The real coefficients \(u_{\mu n}\) are the entries of the orthogonal matrix
\(u\) that diagonalizes \(K\), i.e.
\(
u^{\mathsf T} K u = \mathrm{diag}(\omega_1^{2},\omega_2^{2}),
\)
so that the columns of \(u\) are the normalized eigenvectors of \(K\) appearing
in Eq.~\eqref{eq:A6}
\begin{equation}\label{eq:uevecs}
\mathbf u_n
=
\frac{1}{
\sqrt{
4g^{2}\omega_{c}\omega_{x}
+
\bigl[\omega_{n}^{2}-(\omega_{c}^{2}+4D\omega_{c})\bigr]^{2}
}}
\begin{pmatrix}
2g\sqrt{\omega_{c}\omega_{x}}\\[4pt]
\omega_{n}^{2}-(\omega_{c}^{2}+4D\omega_{c})
\end{pmatrix},
\qquad n=1,2.
\end{equation}
Since \(u\) is orthogonal, the transformation \eqref{eq:A9} preserve canonical relationship for
normal–mode quadratures:
\begin{equation}\label{eq:A10}
[\hat X_n,\hat\Pi_m]=i\delta_{nm}.
\end{equation}
Each normal mode with frequency \(\omega_n\) is then quantized by ladder polariton
operators \(\hat P_n,\hat P_n^\dagger\), defined through
\begin{equation}\label{eq:A11}
\hat X_n=\frac{\hat P_n+\hat P_n^\dagger}{\sqrt{2\omega_n}},
\qquad
\hat\Pi_n=-i\sqrt{\frac{\omega_n}{2}}\,(\hat P_n-\hat P_n^\dagger),
\end{equation}
so that the Hamiltonian becomes
\begin{equation}\label{eq:A12}
\hat H_{\mathrm {HR}}=\sum_{n=1,2}\omega_n\hat P_n^\dagger\hat P_n.
\end{equation}
We now express the bare operators \(\hat a\) and \(\hat b\) in terms of the polaritons.
Using the standard relations between quadratures and ladder operators for the bare modes,
\begin{equation}\label{eq:A13}
\hat a=\sqrt{\frac{\omega_c}{2}}\hat X_a+\frac{i}{\sqrt{2\omega_c}}\hat\Pi_a,
\qquad
\hat b=\sqrt{\frac{\omega_x}{2}}\hat X_b+\frac{i}{\sqrt{2\omega_x}}\hat\Pi_b,
\end{equation}
and inserting the expansions of \(\hat X_\mu\) and \(\hat\Pi_\mu\) with \(\mu = a,b\) in the
normal–mode basis, we obtain
\begin{equation}\label{eq:A14}
\begin{aligned}
\hat X_{\mu} &= \sum_{n}u_{\mu n}\,\hat X_n
        =\sum_{n}u_{\mu n}\frac{\hat P_n+\hat P_n^\dagger}{\sqrt{2\omega_n}},\\[2pt]
\hat\Pi_{\mu} &= \sum_{n}u_{\mu n}\,\hat\Pi_n
        =\sum_{n}u_{\mu n}\Bigl[-i\sqrt{\frac{\omega_n}{2}}\,
           (\hat P_n-\hat P_n^\dagger)\Bigr].
\end{aligned}
\end{equation}
By substituting Eq.~\eqref{eq:A14} into Eq.~\eqref{eq:A13}, one finds
\begin{equation}\label{eq:A15}
\hat a=\sum_{n=1,2}\left(A_n\,\hat P_n+A_n'\,\hat P_n^\dagger\right),
\qquad
\hat b=\sum_{n=1,2}\left(B_n\,\hat P_n+B_n'\,\hat P_n^\dagger\right),
\end{equation}
with Bogoliubov coefficients
\begin{equation}\label{eq:A16}
\begin{aligned}
A_n &= u_{an}\,\frac{\omega_c+\omega_n}{2\sqrt{\omega_c\omega_n}},
&\qquad
A_n' &= u_{an}\,\frac{\omega_c-\omega_n}{2\sqrt{\omega_c\omega_n}},\\[4pt]
B_n &= u_{bn}\,\frac{\omega_x+\omega_n}{2\sqrt{\omega_x\omega_n}},
&\qquad
B_n' &= u_{bn}\,\frac{\omega_x-\omega_n}{2\sqrt{\omega_x\omega_n}}.
\end{aligned}
\end{equation}


\end{document}